\pdfoutput=1
\documentclass[letterpaper, 10 pt, conference]{ieeeconf}  % Comment this line out if you need a4paper
\usepackage{graphicx}      % include this line if your document contains figures
\usepackage{cite}        % required for bibliography
\usepackage{siunitx}
\usepackage{amsmath,cancel}
\usepackage{array}
\usepackage{amsmath}
\usepackage{tikz}
\usepackage{mathdots}
\usepackage{yhmath}
\usepackage{cancel}
\usepackage{color}
\usepackage{multirow}
\usepackage{amssymb}
\usepackage{gensymb}
\usepackage{tabularx}
\usepackage{booktabs}
\usepackage{arydshln}
\usetikzlibrary{fadings}
\usetikzlibrary{patterns}

\newcommand\bovermat[2]{%
	\makebox[0pt][l]{$\smash{\hspace{-10pt}\overbrace{\phantom{%
					\begin{matrix}-\omega _{r\alpha }^{2} & -2{{\beta }_{r}}{{\omega }_{r\alpha }}\end{matrix}}}^{\text{#1}}}$}#2}
\newcommand\undermat[2]{%
	\makebox[0pt][l]{$\smash{\hspace{0pt}\underbrace{\phantom{%
					\begin{matrix}#2\end{matrix}}}_{\text{#1}}}$}#2}

\IEEEoverridecommandlockouts                              % This command is only needed if 
\title{\LARGE \bf
Benefiting from Linear Behaviour of a Nonlinear Reset-based Element at Certain Frequencies
}

\author{Nima Karbasizadeh$^{1}$, Ali Ahmadi Dastjerdi$^{1}$, Niranjan Saikumar$^{1}$, Duarte Val\'erio$^{2}$ and S. Hassan HosseinNia$^{1}$% <-this % stops a space
\thanks{This work was supported by NWO, through OTP TTW project \#16335 and by FCT, through IDMEC, under LAETA, project UID/EMS/50022/2019.}% <-this % stops a space
\thanks{$^{1}$Department of Precision and Microsystem Engineering, Delft University of Technology, Delft, The Netherlands\newline
        {\tt\small \{n.karbasizadehesfahani; a.ahmadidastjerdi; n.saikumar; s.h.hosseinnia\}@tudelft.nl}}%
\thanks{$^{2}$IDMEC, Instituto Superior T\'ecnico, Universidade de Lisboa, Lisbon, Portugal\newline
        {\tt\small  duarte.valerio@tecnico.ulisboa.pt}}%
}
\newtheorem{rem}{Remark}
\newtheorem{thm}{Theorem}
\begin{document}

\maketitle
\thispagestyle{empty}
\pagestyle{empty}

%%%%%%%%%%%%%%%%%%%%%%%%%%%%%%%%%%%%%%%%%%%%%%%%%%%%%%%%%%%%%%%%%%%%%%%%%%%%%%%%
\begin{abstract}

This paper addresses a phenomenon caused by resetting only one of the two states of a so-called second order ``Constant in gain Lead in phase'' (CgLp) element. CgLp is a recently introduced reset-based nonlinear element, bound to circumvent the well-known linear control limitation -- the waterbed effect. The ideal behaviour of such a filter in the frequency domain is unity gain while providing a phase lead for a broad range of frequencies, which clearly violates the linear Bode's gain phase relationship. However, CgLp's ideal behaviour is based on a describing function, which is a first order approximation that neglects the effects of higher order harmonics in the output of the filter. Consequently, achieving the ideal behaviour is challenging when higher order harmonics are relatively large. It is shown in this paper that by resetting only one of the two states of a second order CgLp, the nonlinear filter will act as a linear one at a certain frequency, provided that some conditions are met. This phenomenon can be used to the benefit of reducing higher order harmonics of CgLp's output and achieving the ideal behaviour and thus better performance in terms of precision.

\end{abstract}

%%%%%%%%%%%%%%%%%%%%%%%%%%%%%%%%%%%%%%%%%%%%%%%%%%%%%%%%%%%%%%%%%%%%%%%%%%%%%%%%
\section{Introduction}
Since its formal introduction, dated almost 100 years ago, PID has remained the main control approach used in a wide range of industrial and research applications including precision motion control. However, the increasing demand for faster and at the same time more precise performance has made many researchers to focus on circumventing one severe, fundamental and well-known limitation in linear control theory which is called "waterbed effect", see~\cite{maciejowski1989multivariable}.  Referring to frequency loop-shaping method for designing a controller, one can understand that increasing the gain of open loop frequency response at lower frequencies and decreasing it at higher frequencies will result in better performance in terms of tracking and steady state precision, see~\cite{schmidt2014design}. However, Bode's gain-phase relationship for linear systems, along with the frequency response of the differentiator of PID, will bring the desire for precision to a contradiction with the robustness of the system. Among all the efforts made to get around this limitation using nonlinearity, a category of researches are based on introducing a relatively simple nonlinearity to system, namely reset technique, see~\cite{clegg1958nonlinear, horowitz1975non}. \\
Reset control is based on the idea of resetting the states of the controller, provided that the resetting condition is met. The concept was firstly shown in~\cite{clegg1958nonlinear}, in which a nonlinear reset integrator, thereafter called Clegg integrator, demonstrated significantly less phase lag than a linear one while maintaining the gain behaviour according to describing function approximation. The idea has then been further developed to create more sophisticated reset elements such as First Order Reset Element (FORE) in~\cite{horowitz1975non, zaccarian2005first}, Generalized FORE (GFORE) in~\cite{guo2009frequency} and Second Order Reset Element (SORE) in~\cite{hazeleger2016second}. Researchers took advantage of the reset elements in different capacities such as phase lag reduction, decreasing sensitivity peak, narrowband and broadband phase compensation, see~\cite{wu2006reset, li2005nonlinear,li2010reset,li2011optimal,palanikumar2018no,saikumar2019resetting}.\\
A recent research has used FORE and SORE in combination with a linear lead to create a filter which has constant gain while producing a phase lead in a broad range of frequencies~\cite{saikumar2019constant}. The so-called ``Constant in Gain Lead in Phase'' (CgLp) can be used in the framework of PID, completely replacing or taking up a big portion of derivative duties, which is providing the required phase lead in the bandwidth region for the system to be robustly stable. Unlike the derivative in PID, CgLp does not violate the loop-shaping requirement. However, achieving the desired ideal behaviour of CgLp can be challenging when the higher order harmonics of its output are relatively large, since the ideal behaviour is based on the assumptions of the describing function method. This is a first order approximation, and thus the effects of higher order harmonics are neglected.\\
This paper will introduce and investigate a phenomenon that can happen in a CgLp designed based on SORE. So far, in all of the researches done on SORE, both states of such a filter were reset with same resetting factor. But what happens if one resets only one state of a second order reset element? This paper will show that under certain conditions, a SORE which only has one resetting state will behave like a linear filter at a certain frequency. Hence, the higher order harmonics will be zero at that frequency and the element will have the ideal behaviour defined by describing function.\\
The remainder of this paper is organized as follows. The second section presents the preliminaries. The following one introduces and studies the case in which only one state of a SORE in CgLp framework is being reset. The third section will investigate the benefits and applications of the interesting phenomenon in presented CgLp. Finally, the paper concludes with some remarks and recommendations about ongoing works.

\section{Preliminaries}

In this section, the preliminaries of this study will be discussed.

\subsection{General Reset Controller}
A general form of a reset controller is as follows~\cite{Guo:2015}:
\begin{align}
\label{eq:reset}
{{\sum }_{R}}:=\left\{ \begin{aligned}
& {{{\dot{x}}}_{r}}(t)={{A}}{{x}_{r}}(t)+{{B}}e(t)&\text{if }e(t)\ne 0,\\ 
& {{x}_{r}}({{t}^{+}})={{A}_{\rho }}{{x}_{r}}(t)&\text{if }e(t)=0, \\ 
& u(t)={{C}}{{x}_{r}}(t)+{{D}}e(t) \\ 
\end{aligned} \right.
\end{align}
where $A,B,C,D$ are the state space matrices of the base linear system and $A_\rho=\text{diag}(\gamma_1,...,\gamma_n)$ is called reset matrix. This contains the reset coefficients for each state which are denoted by $\gamma_1,...,\gamma_n$. The controller's input and output are represented by $e(t)$ and $ u(t) $, respectively.
\subsection{Describing Functions}
Like many other nonlinear controllers, the steady state response of a reset element to a sinusoidal input is not sinusoidal. Thus, its frequency response has been analysed by Describing Function (DF) method in the literature~\cite{guo2009frequency}. However, the DF method only takes the first harmonic of Fourier series decomposition of the output into account and neglects the effects of the higher order harmonics. As it will be shown in this paper, this simplification can sometimes be significantly inaccurate. To have more accurate  information about the frequency response of nonlinear systems, a method called ``Higher Order Sinusoidal Input Describing Function'' (HOSIDF) has been introduced in~\cite{nuij2006higher}. In~\cite{kars2018HOSIDF,dastjerdi2020closed} the HOSIDF has been developed for reset elements defined by~(\ref{eq:reset}) as follows:
\begin{align}  \nonumber
& G_n(\omega)=\left\{ \begin{aligned}
& C{{(j\omega I-A)}^{-1}}(I+j{{\Theta }_{D}}(\omega ))B+D\quad n=1\\ 
& C{{(j\omega nI-A)}^{-1}}j{{\Theta }_{D}}(\omega )B\qquad\quad\text{odd }n> 2\\ 
& 0\quad\qquad\qquad\qquad\qquad\qquad\qquad\text{ even }n\ge 2\\ 
\end{aligned} \right. \\
&\begin{aligned}
& {{\Theta }_{D}}(\omega )=-\frac{2{{\omega }^{2}}}{\pi }\Delta (\omega )[{{\Gamma }_{r}}(\omega )-{{\Lambda }^{-1}}(\omega )] \\  
& \Lambda (\omega )={{\omega }^{2}}I+{{A}^{2}} \\  
& \Delta (\omega )=I+{{e}^{\frac{\pi }{\omega }A}} \\  
& {{\Delta }_{r}}(\omega )=I+{{A}_{\rho}}{{e}^{\frac{\pi }{\omega }A}} \\  
& {{\Gamma }_{r}}(\omega )={{\Delta }_{r}}^{-1}(\omega ){{A}_{\rho}}\Delta (\omega ){{\Lambda }^{-1}}(\omega ) \\
\end{aligned} 
\end{align}
where $G_n(\omega)$ is the $n^{\text{th}}$ harmonic describing function for sinusoidal input with frequency of $\omega$.
\subsection{CgLp}
According to~\cite{saikumar2019constant}, CgLp is a broadband phase compensation element whose first harmonic gain behaviour is constant while providing a phase lead. Two architectures for CgLp are suggested using FORE or SORE, both consisting in a reset lag element in series with a linear lead filter, namely $R$ and $D$. For FORE CgLp:
\begin{align}
\label{eq:fore}
&R(s)=\cancelto{A_\rho}{\frac{1}{{s}/{{{\omega }_{r\alpha }}+1}\;}},&D(s)=\frac{{s}/{{{\omega }_{r}}}\;+1}{{s}/{{{\omega }_{f}}}\;+1}
\end{align}
For SORE CgLp:
\begin{equation}
\label{eq:sore}
\begin{aligned}
&R(s)=\cancelto{A_\rho}{\frac{1}{({s}/{{{\omega }_{r\alpha }}{{)}^{2}}+(2s{{{\beta }_{r}}}/{{{\omega }_{r\alpha }})}\;+1}\;}}\\ \\
&D(s)=\frac{({s}/{{{\omega }_{r}}{{)}^{2}}+(2s{{{\beta }_{r}}}/{{{\omega }_{r}})}\;+1}\;}{({s}/{{{\omega }_{f}}{{)}^{2}}+(2s{{}}/{{{\omega }_{f}})}\;+1}\;}	
\end{aligned}
\end{equation}
In~(\ref{eq:fore}) and~(\ref{eq:sore}), $\omega_{r\alpha}=\omega_r/\alpha$, $\alpha$ is a tuning parameter accounting for a shift in corner frequency of the filter due to resetting action, $\beta_{r}$ is the damping coefficient and $[\omega_{r},\omega_{f}]$ is the frequency range where the CgLp will provide the required phase lead. The arrow indicates that the states of element are reset according to $A_\rho$; i.e. are multiplied by $A_\rho$ when the reset condition is met.
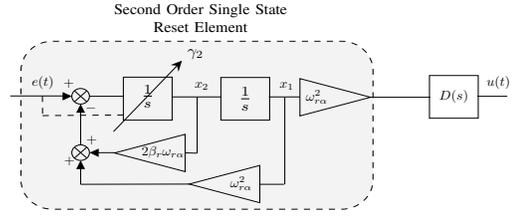
\begin{figure}[t!]
	\centering
	\scalebox{0.8}[0.75]{
		\resizebox{\columnwidth}{!}{

			\tikzset{every picture/.style={line width=0.75pt}} %fault line width to 0.75pt        
			
			\begin{tikzpicture}[x=0.75pt,y=0.75pt,yscale=-1,xscale=1]
			%uncomment if require: \path (0,231.3333282470703); %set diagram left start at 0, and has height of 231.3333282470703
			
			%Flowchart: Alternative Process [id:dp6881341252372146] 
			\draw  [fill={rgb, 255:red, 0; green, 0; blue, 0 }  ,fill opacity=0.05 ][dash pattern={on 4.5pt off 4.5pt}] (21.81,81.3) .. controls (21.81,64.6) and (35.35,51.06) .. (52.05,51.06) -- (330.86,51.06) .. controls (347.56,51.06) and (361.1,64.6) .. (361.1,81.3) -- (361.1,193.61) .. controls (361.1,210.31) and (347.56,223.85) .. (330.86,223.85) -- (52.05,223.85) .. controls (35.35,223.85) and (21.81,210.31) .. (21.81,193.61) -- cycle ;
			%Shape: Rectangle [id:dp9546542540773222] 
			\draw   (120.5,88) -- (167.5,88) -- (167.5,132.22) -- (120.5,132.22) -- cycle ;
			%Flowchart: Summing Junction [id:dp6216795659062622] 
			\draw   (70.83,107.67) .. controls (70.83,103.25) and (74.64,99.67) .. (79.33,99.67) .. controls (84.03,99.67) and (87.83,103.25) .. (87.83,107.67) .. controls (87.83,112.08) and (84.03,115.67) .. (79.33,115.67) .. controls (74.64,115.67) and (70.83,112.08) .. (70.83,107.67) -- cycle ; \draw   (73.32,102.01) -- (85.34,113.32) ; \draw   (85.34,102.01) -- (73.32,113.32) ;
			%Straight Lines [id:da5395288542551473] 
			\draw    (11.5,107.33)  -- (68.83,107.33) ;
			\draw [shift={(70.83,107.67)}, rotate = 180.36] [fill={rgb, 255:red, 0; green, 0; blue, 0 }  ][line width=0.75]  [draw opacity=0] (8.93,-4.29) -- (0,0) -- (8.93,4.29) -- cycle    ;
			
			%Straight Lines [id:da0249709012696957] 
			\draw    (87.83,107.67)-- (120.67,107.67) ;

			%Straight Lines [id:da15844708457609946] 
			\draw    (111,144) -- (175.66,72.48) ;
			\draw [shift={(177,71)}, rotate = 492.12] [fill={rgb, 255:red, 0; green, 0; blue, 0 }  ][line width=0.75]  [draw opacity=0] (10.72,-5.15) -- (0,0) -- (10.72,5.15) -- (7.12,0) -- cycle    ;
			
			%Shape: Rectangle [id:dp7988875530096091] 
			\draw   (214.5,88.33) -- (261.5,88.33) -- (261.5,132.22) -- (214.5,132.22) -- cycle ;
			%Straight Lines [id:da6934013984849898] 
			\draw    (167.83,108.33) -- (214.67,108.33) ;

			%Flowchart: Summing Junction [id:dp7781755232785939] 
			\draw   (70.83,165.67) .. controls (70.83,161.25) and (74.64,157.67) .. (79.33,157.67) .. controls (84.03,157.67) and (87.83,161.25) .. (87.83,165.67) .. controls (87.83,170.08) and (84.03,173.67) .. (79.33,173.67) .. controls (74.64,173.67) and (70.83,170.08) .. (70.83,165.67) -- cycle ; \draw   (73.32,160.01) -- (85.34,171.32) ; \draw   (85.34,160.01) -- (73.32,171.32) ;
			%Straight Lines [id:da0709925984760833] 
			\draw    (79.33,117.67) -- (79.33,157.67) ;
			
			\draw [shift={(79.33,115.67)}, rotate = 90] [fill={rgb, 255:red, 0; green, 0; blue, 0 }  ][line width=0.75]  [draw opacity=0] (8.93,-4.29) -- (0,0) -- (8.93,4.29) -- cycle    ;
			%Straight Lines [id:da8013227443725106] 
			\draw    (89.83,165.69) -- (112,165.69) ;
			
			\draw [shift={(87.83,165.67)}, rotate = 0.79] [fill={rgb, 255:red, 0; green, 0; blue, 0 }  ][line width=0.75]  [draw opacity=0] (8.93,-4.29) -- (0,0) -- (8.93,4.29) -- cycle    ;
			%Flowchart: Extract [id:dp2873122281307685] 
			\draw   (112,166) -- (180,146.81) -- (180,185.19) -- cycle ;
			%Straight Lines [id:da947321607603828] 
			\draw    (191.58,108.69) -- (191.58,166.22) ;

			%Straight Lines [id:da21171374588013614] 
			\draw    (180,165.89) -- (191.33,166.22) ;

			%Straight Lines [id:da9529150432534255] 
			\draw    (260.83,108.33) -- (290.67,108.33) ;

			%Flowchart: Extract [id:dp463175866425624] 
			\draw   (358.67,108.67) -- (290.67,89.47) -- (290.67,127.86) -- cycle ;
			%Flowchart: Extract [id:dp719939512345652] 
			\draw   (184,198) -- (252,178.81) -- (252,217.19) -- cycle ;
			%Straight Lines [id:da22602627890287041] 
			\draw    (276.08,108.69) -- (276.08,198.22) ;

			%Straight Lines [id:da09763427765450494] 
			\draw    (252,198.22) -- (275.33,198.22) ;

			%Straight Lines [id:da3275341587354579] 
			\draw    (79.33,198.22) -- (184,198) ;

			%Straight Lines [id:da038281499213885795] 
			\draw    (79.33,175.67) -- (79.33,198.22) ;
			
			\draw [shift={(79.33,173.67)}, rotate = 90] [fill={rgb, 255:red, 0; green, 0; blue, 0 }  ][line width=0.75]  [draw opacity=0] (8.93,-4.29) -- (0,0) -- (8.93,4.29) -- cycle    ;
			%Straight Lines [id:da5890933505833742] 
			\draw  [dash pattern={on 4.5pt off 4.5pt}]  (42.49,107.49) -- (42.67,127.56) ;

			%Straight Lines [id:da3975047933019873] 
			\draw  [dash pattern={on 4.5pt off 4.5pt}]  (42.67,127.56) -- (120.67,126.89) ;

			%Shape: Rectangle [id:dp6020970624218196] 
			\draw   (415.5,86.33) -- (462.5,86.33) -- (462.5,130.22) -- (415.5,130.22) -- cycle ;
			%Straight Lines [id:da8834054941716238] 
			\draw    (358.67,108.67) -- (416.5,108.67) ;

			%Straight Lines [id:da3931800790814295] 
			\draw    (461.67,107.67) -- (490.5,107.67) ;

			% Text Node
			\draw (144,107.5) node [scale=1.5]   {$\frac{1}{s}$};
			% Text Node
			\draw (43,94) node   {$e( t)$};
			% Text Node
			\draw (196,97) node   {$x_{2}$};
			% Text Node
			\draw (190,64) node  [scale=1.2] {$\gamma _{2}$};
			% Text Node
			\draw (238,110.28) node  [scale=1.5] {$\frac{1}{s}$};
			% Text Node
			\draw (278,97) node   {$x_{1}$};
			% Text Node
			\draw (158.67,164.67) node   {$2\beta_r \omega _{r\alpha }$};
			% Text Node
			\draw (307,108) node   {$\omega ^{2}_{r\alpha }$};
			% Text Node
			\draw (234,197.33) node   {$\omega ^{2}_{r\alpha }$};
			% Text Node
			\draw (68,94.67) node   {$+$};
			% Text Node
			\draw (89.33,120) node   {$-$};
			% Text Node
			\draw (90,153) node   {$+$};
			% Text Node
			\draw (68.67,174.67) node   {$+$};
			% Text Node
			\draw (439,108.28) node   {$D( s)$};
			% Text Node
			\draw (391,95) node   {};
			% Text Node
			\draw (482,93) node   {$u( t)$};
			% Text Node
			\draw (195,27) node  [align=center] [scale=1.15] {Second Order Single State \\ Reset Element};

			\end{tikzpicture}
		}
	}
	\caption{Block Diagram of a SOSRE CgLp. The second integrator is not being reset which translates to $\gamma_1=~1$.}
	\label{fig:GSORE_blockdiag}
\end{figure}

\section{Second Order Single State Reset Element}
This section addresses the architecture and frequency behaviour of a Second Order Single State Reset Element (SOSRE), in framework of CgLp. SOSRE is in fact a special case of a  with only one resetting state.
\subsection{Architecture and State Space Representation}
%There are multiple state space realizations for a SORE in a CgLp, e.g., controllable, observable and Jordan canonical forms. One can notice that while the reset coefficients (i.e., $A_\rho$ diagonal elements) are equal, the realization form of SORE will not affect the output and describing functions of the frequency response. Nevertheless, changing the form will change the describing functions, especially the higher order ones, when  using different resetting coefficients, $\gamma_i$. This is due to the fact that in each form of realization, states are defined differently. In other words, among different state space forms, states contain different information about the system; thus, resetting them with an identical $A_\rho$ will have non-identical outcomes. Obviously, in the special case of equal resetting factor, the output equation of state space representation will finally build up the same output, irrespective of the form. \\
Figure~\ref{fig:GSORE_blockdiag} shows the block diagram of the SOSRE. The architecture is similar to SORE in controllable canonical form with the difference being that the second integrator \textemdash the first state in controllable canonical state space realization, $x_1$, is not being reset, i.e., $\gamma_1=1$. This specific type of resetting in which a resetting state and a non-resetting one are coupled creates an interesting behaviour for this element in terms of steady state output. State space representation of SOSRE in the framework of CgLp is: 
\begin{equation}
\begin{aligned}
\\
& A=\left[ \, {\begin{array}{cc;{2pt/1.5pt}cc}
\bovermat{SOSRE}{0 & 1} & 0 & 0  \\
-\omega _{r\alpha }^{2} & -2{{\beta }_{r}}{{\omega }_{r\alpha }} & 0 & 0  \\ \hdashline[2pt/1.5pt]
0 & 0 & 0 & 1  \\
\omega _{r\alpha }^{2} & 0 & \undermat{Second order lead}{-\omega _{f}^{2} & -2{{\omega }_{f}}}  \\
\end{array}} \right],B=\left[ \begin{matrix}
0  \\
1  \\
0  \\
0  \\
\end{matrix} \right], \\ \\
& C=\left[ \begin{matrix}
{{\left( \frac{{{\omega }_{r\alpha }}{{\omega }_{f}}}{{{\omega }_{r}}} \right)}^{2}} & 0 & \omega _{f}^{2}\left( 1-{{\left( \frac{{{\omega }_{f}}}{{{\omega }_{r}}} \right)}^{2}} \right) & \omega _{f}^{2}\left( \frac{2{{\beta }_{r}}}{{{\omega }_{r}}}-\frac{2{{\omega }_{f}}}{\omega _{r}^{2}} \right)  \\
\end{matrix} \right], \\ 
& D=\left[ 0 \right], {{A}_{\rho }}=\text{diag}(1,\gamma_2,1,1).
\end{aligned}
\end{equation} 
It has to be mentioned that since SOSRE is a nonlinear element, transforming the above state space representation to other forms may result in a different behaviour of the element. In other words, state space representation should exactly match the block diagram represented in Fig.~\ref{fig:GSORE_blockdiag}.
\begin{rem}
	\label{remark}	
	Assuming a sinusoidal input to a reset element, if the phase shift between the output of its base linear system and its input is zero, the reset action will be of no effect in steady state response, and thus the reset element can be regarded as a linear system in terms of steady state response at that certain frequency.
\end{rem}
\begin{figure}[t!]
	\centering
	\scalebox{1}[0.9]{
		\includegraphics[width=\columnwidth]{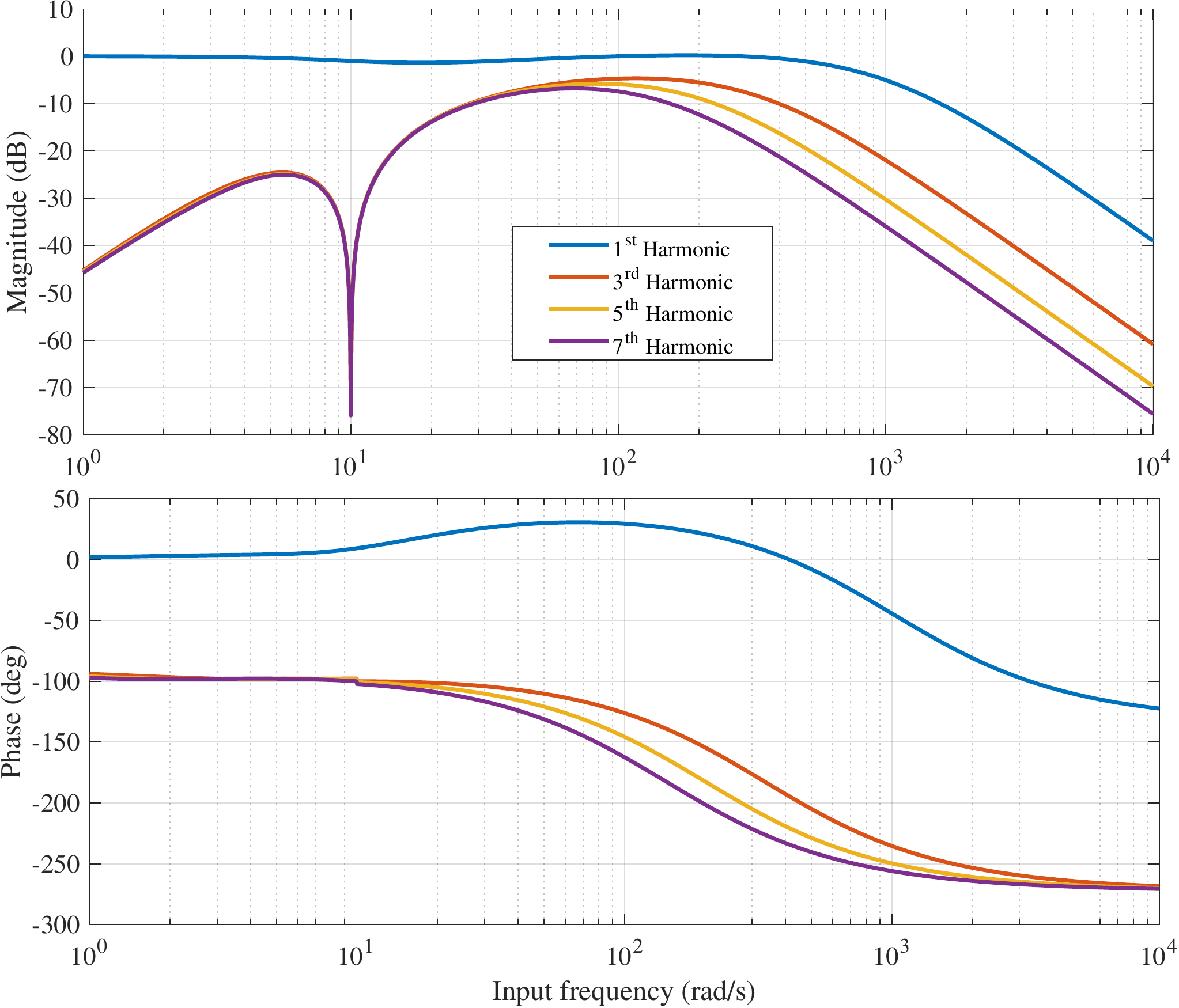}
	}
	\caption{Harmonics of SOSRE CgLp utilizing HOSIDF method.}
	\label{fig:hosidf_SORE}
\end{figure}
The proof of this is trivial, since the reset element under such circumstances will reset its output when its output is already at zero, resulting in no change from the resetting action. In the case of a SOSRE, if $e(t)=\sin(\omega t)$, the reset action of the first integrator will be of no effect if:
\begin{align}
\begin{aligned}
& \angle \frac{{{X}_{1}}(j\omega )}{E(j\omega )}=\frac{\pi}{2}-{{\tan }^{-1}}\left(\frac{2{{\beta }_{r}}{{\omega }_{r\alpha }}\omega }{-{{\omega }^{2}}+{{\omega }_{r\alpha }^{2}}}\right)=0 \\ 
& \Rightarrow \omega ={{\omega }_{r\alpha }}.  
\end{aligned}
\end{align}
Since there is no other nonlinear element in SOSRE, it will behave like a linear element at frequency $\omega_{r\alpha}$. Solving such an equation for a conventional FORE will result in $\omega=0$ as the only solution and thus it does not exhibit such a behaviour.\\

\subsection{HOSIDF of SOSRE CgLp}
%Assume a state-space representation of a SOSRE CgLp system:
%\begin{align}
%\label{eq:partial_SORE}
%\begin{aligned}
%&A=\left[ \begin{matrix}
%0 & 1 & 0 & 0  \\
%-100 & -20 & 0 & 0  \\
%0 & 0 & 0 & 1  \\
%100 & 0 & -1e6 & -2000  \\
%\end{matrix} \right],B=\left[ \begin{matrix}
%0  \\
%1  \\
%0  \\
%0  \\
%\end{matrix} \right],\\
%&C=\left[ \begin{matrix}
%7.831e5 & 0 & -7.83e9 & -1.549e7  \\
%\end{matrix} \right],D=0,\\
%&{{A}_{\rho }}=\text{diag}(1,0.1,1,1)
%\end{aligned}
%\end{align}
Assume a state-space representation of a SOSRE CgLp system with the following configuration
\begin{align}
\label{eq:partial_SORE}
\begin{aligned}
\omega_{r\alpha}=10&, \beta_{r}=1, \alpha=1.13, \omega_{f}=1000\\
{{A}_{\rho }}&=\text{diag}(1,0.1,1,1)
\end{aligned}
\end{align}
\begin{figure}[t!]
	\centering
	\scalebox{1}[0.9]{
		\includegraphics[width=\columnwidth]{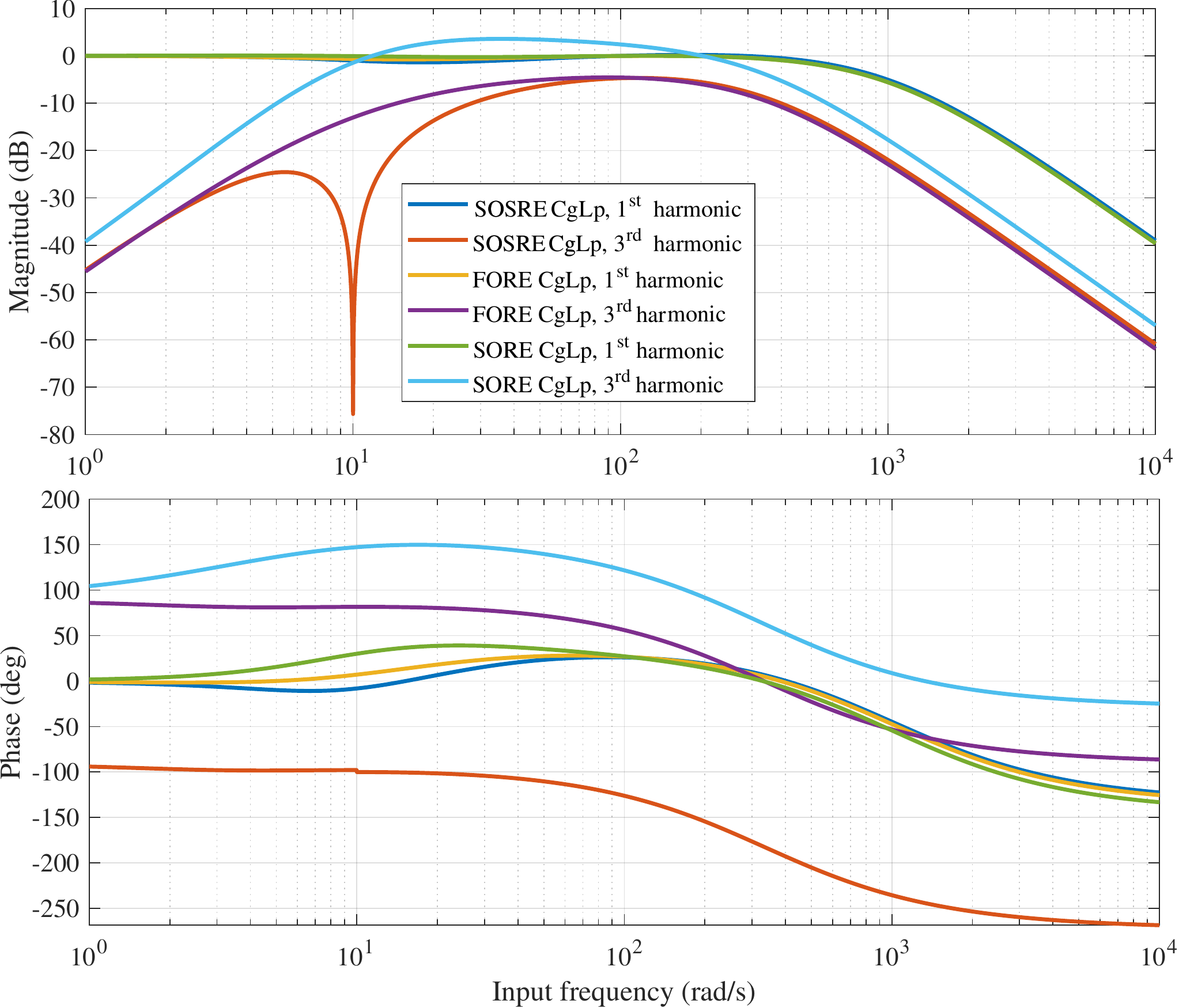}
	}
	\caption{Comparison of the first and third harmonics of conventional SORE CgLp in which $ \gamma_1=\gamma_2=0.44 $, SOSRE CgLp in which $ \gamma_1=1 ,\gamma_2=0.1 $ and  FORE CgLp in which $ \gamma=0.15 $. In all CgLps $\omega_{r\alpha}=10 $ and frequency range is $ [10,1000] $ rad/s and in SORE and SOSRE, $ \beta_r=1 $. All CgLps designed to have matching $1^{st}$ order harmonic gain. They are also designed to have the same first order harmonic phase at $100$ rad/s.}
	\label{fig:compare_hosidf}
\end{figure}\\
Figure~\ref{fig:hosidf_SORE} depicts the $1^{st}, 3^{rd}, 5^{th}$ and $7^{th}$  order describing functions of this CgLp in frequency domain. It goes without saying that the steady state output of a linear system, when the input is sinusoidal, is also a sinusoid, and can consequently be completely described by the first term of a Fourier series; thus, all higher order harmonics are zero. Figure~\ref{fig:hosidf_SORE} shows that this is the case for SOSRE CgLp as well, at $\omega=10$~rad/s, where its behaviour is like that of a linear filter. The benefits of this phenomenon will be discussed in following sections.
%\begin{figure}[t!]
%	\centering
%	\includegraphics[width=\columnwidth]{compare_hosidf.pdf}
%	\caption{Comparison of the first and third harmonic  of conventional SORE CgLp in which $ \gamma_1=\gamma_2=\gamma $ and SOSRE CgLp in which $ \gamma_1=1 ,\gamma_2=\gamma $ and  FORE CgLp. In all CgLps $\omega_{r\alpha}=10 $ and frequency range is $ [10,1000] $ and in SORE ones, $ \beta_r=1 $.}
%	\label{fig:compare_hosidf}
%\end{figure}
\subsection{Comparison with FORE CgLp and Conventional SORE CgLp}
\label{sec:comparison}
SOSRE CgLp is a special case of a general SORE CgLp in which only one state is being reset. The fundamental distinction of SOSRE with respect to SORE and FORE is that it has a reset state and a non-reset state that are coupled together, which is in fact, the main reason for the linear behaviour. However, using these three elements in framework of CgLp, one can achieve the same gain behaviour in DF for all three, while different higher order harmonic behaviour. Figure~\ref{fig:compare_hosidf} compares DF and HOSIDF of the SOSRE CgLp described in~(\ref{eq:partial_SORE}) with a conventional SORE CgLp realized in controllable form with the following configuration:
\begin{align}
\label{eq:conventional_SORE}
\begin{aligned}\omega_{r\alpha}=10&, \beta_{r}=1, \alpha=0.9,  \omega_{f}=1000\\
{{A}_{\rho }}&=\text{diag}(0.44,0.44,1,1)
\end{aligned}
\end{align}
The comparison also includes a FORE CgLp with the following configuration:
%\begin{align}
%\label{eq:conventional_SORE}
%\begin{aligned}
%&A=\left[ \begin{matrix}
%0 & 1 & 0 & 0  \\
%-100 & -20 & 0 & 0  \\
%0 & 0 & 0 & 1  \\
%100 & 0 & -1e6 & -2000  \\
%\end{matrix} \right],B=\left[ \begin{matrix}
%0  \\
%1  \\
%0  \\
%0  \\
%\end{matrix} \right],\\
%&C=\left[ \begin{matrix}
%1.235e6 & 0 & -1.234e10 & -2.447e7  \\
%\end{matrix} \right],D=0,\\
%&{{A}_{\rho }}=\text{diag}(0.44,0.44,1,1)
%\end{aligned}
%\end{align}
%\quad where SORE is realized in controllable form with parameters, $\omega_{r\alpha}=10, \beta_{r}=1, \alpha=0.9$ and $\omega_{f}=1000$. The comparison also includes a FORE CgLp as:
%\begin{align}
%\label{eq:FORE}
%\begin{aligned}
%&A=\left[ \begin{matrix}
%	-10 & 0 & 0  \\
%	0 & 0 & 1  \\
%	10 & -1e6 & -2000  \\
%\end{matrix} \right],B=\left[ \begin{matrix}
%	1  \\
%	0  \\
%	0  \\
%\end{matrix} \right],\\
%&C=\left[ \begin{matrix}
%	0 & 1e6 & 7.692e4  \\
%\end{matrix} \right],D=0,\\
%&{{A}_{\rho }}=\text{diag}(0.15,1,1)
%\end{aligned}
%\end{align}
\begin{align}
\label{eq:FORE}
\begin{aligned}
\omega_{r\alpha}&=10, \alpha=1.3, \omega_{f}=1000\\
{{A}_{\rho }}&=\text{diag}(0.15,1,1)
\end{aligned}
\end{align}
In order to make all CgLps behave the same at high frequencies, an additional low-pass filter has been added to FORE CgLp with the same corner frequency of $1000~\text{rad}/{\text{s}}$. Moreover, the $\gamma_i$ values are chosen in such a manner that all filters have the same $1^{st}$ harmonic phase at $100~\text{rad}/{\text{s}}$ and $\alpha$ values are chosen for all CgLps to have unity gain at $100~\text{rad}/{\text{s}}$.\\
According to Fig.~\ref{fig:compare_hosidf}, although all three CgLps have almost the same first order gain behaviour, the third harmonic is quite different. FORE and SOSRE have the same behaviour for the harmonics except for the range of $ [3,70]~\text{rad}/\text{s} $ where SOSRE CgLp has considerably smaller third order harmonic. The conventional SORE CgLp has noticeably larger $3^\text{rd}$ order harmonic, to the extent that it dominates the first harmonic in a large range of frequencies.\\
It should be noted that other higher order harmonics, i.e., $5^\text{th}, 7^\text{th},$ etc. will follow the same trend and as seen in Fig.~\ref{fig:hosidf_SORE} are descending in magnitude with respect to their order; however, they are not depicted in Fig.~\ref{fig:compare_hosidf} for the sake of plot clarity.
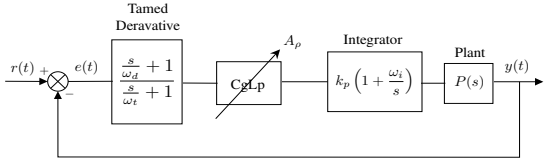
\begin{figure}[t!]
	\centering
	\scalebox{0.85}[0.8]{
		\resizebox{\columnwidth}{!}{
			
			\tikzset{every picture/.style={line width=0.75pt}} %set default line width to 0.75pt        
			
			\begin{tikzpicture}[x=0.75pt,y=0.75pt,yscale=-1,xscale=1]
			%uncomment if require: \path (0,203); %set diagram left start at 0, and has height of 203
			
			%Shape: Rectangle [id:dp6004374397694128] 
			\draw  [line width=0.75]  (113.21,49.25) -- (188.5,49.25) -- (188.5,145) -- (113.21,145) -- cycle ;
			%Shape: Rectangle [id:dp9604039091965582] 
			\draw  [line width=0.75]  (225.5,74.17) -- (294.5,74.17) -- (294.5,123) -- (225.5,123) -- cycle ;
			%Shape: Rectangle [id:dp5038121848240322] 
			\draw  [line width=0.75]  (344.5,63) -- (445.5,63) -- (445.5,131) -- (344.5,131) -- cycle ;
			%Straight Lines [id:da7731408703992304] 
			\draw [line width=0.75]    (66.42,96.67) -- (114.22,96.45) ;

			%Straight Lines [id:da36461852787237525] 
			\draw [line width=0.75]    (550,97) -- (550,183) -- (55.5,182.67) -- (55.46,109.73) ;
			\draw [shift={(55.46,107.73)}, rotate = 449.97] [fill={rgb, 255:red, 0; green, 0; blue, 0 }  ][line width=0.75]  [draw opacity=0] (8.93,-4.29) -- (0,0) -- (8.93,4.29) -- cycle    ;
			
			%Straight Lines [id:da1393811703303558] 
			\draw [line width=0.75]    (-1,96.45) -- (42.5,96.66) ;
			\draw [shift={(44.5,96.67)}, rotate = 180.27] [fill={rgb, 255:red, 0; green, 0; blue, 0 }  ][line width=0.75]  [draw opacity=0] (8.93,-4.29) -- (0,0) -- (8.93,4.29) -- cycle    ;
			
			%Straight Lines [id:da4183844429842236] 
			\draw [line width=0.75]    (522.5,97) -- (573.5,97) ;
			\draw [shift={(575.5,97)}, rotate = 180] [fill={rgb, 255:red, 0; green, 0; blue, 0 }  ][line width=0.75]  [draw opacity=0] (8.93,-4.29) -- (0,0) -- (8.93,4.29) -- cycle    ;
			
			%Straight Lines [id:da6169379750986348] 
			\draw [line width=0.75]    (189.42,98.45) -- (226.5,98.67) ;

			%Straight Lines [id:da8911932794469448] 
			\draw [line width=0.75]    (295.5,97) -- (344.5,97) ;

			%Shape: Rectangle [id:dp780701248947808] 
			\draw  [line width=0.75]  (469.5,76.17) -- (521.5,76.17) -- (521.5,121.67) -- (469.5,121.67) -- cycle ;
			%Straight Lines [id:da03234018932258742] 
			\draw [line width=0.75]    (444.5,98.67) -- (470.5,98.67) ;

			%Straight Lines [id:da465750780248396] 
			\draw [line width=0.75]    (224.5,139.08) -- (292.18,61.59) ;
			\draw [shift={(293.5,60.08)}, rotate = 491.13] [fill={rgb, 255:red, 0; green, 0; blue, 0 }  ][line width=0.75]  [draw opacity=0] (10.72,-5.15) -- (0,0) -- (10.72,5.15) -- (7.12,0) -- cycle    ;
			
			%Flowchart: Summing Junction [id:dp09605751187284017] 
			\draw   (44.5,96.67) .. controls (44.5,90.56) and (49.41,85.61) .. (55.46,85.61) .. controls (61.52,85.61) and (66.42,90.56) .. (66.42,96.67) .. controls (66.42,102.77) and (61.52,107.73) .. (55.46,107.73) .. controls (49.41,107.73) and (44.5,102.77) .. (44.5,96.67) -- cycle ; \draw   (47.71,88.85) -- (63.21,104.49) ; \draw   (63.21,88.85) -- (47.71,104.49) ;
			
			% Text Node
			\draw (395,97) node [scale=1.2]  {$k_{p}\left( 1+\dfrac{\omega _{i}}{s}\right)$};
			% Text Node
			\draw (307,51.91) node [scale=1.2]  {$A_{\rho }$};
			% Text Node
			\draw (17.84,81.76) node [scale=1.2]  {$r( t)$};
			% Text Node
			\draw (547.62,79.76) node [scale=1.2]  {$y( t)$};
			% Text Node
			\draw (86.35,80.76) node [scale=1.2]  {$e( t)$};
			% Text Node
			\draw (150.85,97.13) node [scale=1.5]  {$\dfrac{\frac{s}{\omega _{d}} +1}{\frac{s}{\omega _{t}} +1}$};
			% Text Node
			\draw (496.57,97.58) node [scale=1.2]  {$P( s)$};
			% Text Node
			\draw (152,23) node [scale=1.2] [align=center] {Tamed \\ Deravative};
			% Text Node
			\draw (393,50) node [scale=1.2] [align=center] {Integrator};
			% Text Node
			\draw (497,63) node [scale=1.2] [align=center] {Plant};
			% Text Node
			\draw (259,102) node [scale=1.2]  [align=left] {\text{CgLp}};
			% Text Node
			\draw (41,85) node   {$+$};
			% Text Node
			\draw (68,111) node   {$-$};

			\end{tikzpicture}
		}
	}
	\caption{Designed control architecture to compare the performance of CgLps presented in~(\ref{eq:partial_SORE}),~(\ref{eq:conventional_SORE}) and~(\ref{eq:FORE}) to control the plant introduced in~(\ref{eq:plant}).} 
	\label{fig:openloop_blockdiagram}
\end{figure} 
\section{On benefits of the notch-like behaviour in higher order harmonics of SOSRE CgLp}
As aforementioned, designing a controller in frequency domain is a very popular method. However, since no method exists for capturing all the frequency aspects of a nonlinear reset element, DF approximation is being used for frequency domain design. But how reliable is this approximation? The approximation is based on the assumption that the first harmonic of the steady state response is the dominant one and higher order harmonics are negligible. And since the first harmonic gain is dominant, the phase behaviour of the controller will follow the first harmonic phase. It can be concluded that smaller the higher order harmonics are, closer the real controller is to its design based on DF.\\% Therefore, one can claim that decreasing the higher order harmonics while maintaining the desired first order one can be an objective in designing reset controllers.\\
However, referring to the example comparison made in Section~\ref{sec:comparison}, this assumption is not true for all the cases and not only is the approximation not accurate, but also it can be completely misleading in some cases like conventional SORE CgLp presented in~(\ref{eq:conventional_SORE}); where, in a wide range of the frequencies, the third harmonic dominates the first one and thus the DF and the design based on it are completely unreliable.  Although this degenerate case was readily observable for conventional SORE CgLp in HOSIDF of the controller itself, in some cases, it can only be seen in HOSIDF of the overall open loop system including the plant, due to a phenomenon mentioned in~\cite{kars2018HOSIDF}. HOSIDF of the open loop can be obtained as follows:
\begin{equation}
L_n(\omega)=G_n(\omega)C(n\omega)P(n\omega)
\end{equation}
where $C(\omega)$ is the DF of the linear part of the controller and $P(\omega)$ is the DF of the plant. The above equation reveals that in open loop frequency response of a nonlinear controller together with a mass-spring-damper system which has a resonance at $\omega_n$, the resonance peak for the third harmonic will happen at $\omega_n/3$, the peak for the fifth at $\omega_n/5$, and so on. Consequently, if the controller happens to have a large enough third order harmonic even if it is not readily dominating the first one, the resonance peak can cause it to dominate. However, according to the notch-like HOSIDF behaviour of a SOSRE, this controller can be designed in a manner to cancel the third order harmonic resonance peak. For such a purpose, $\omega_{r\alpha }$ should be designed to be equal $\omega_n/3$.
\begin{figure}[t!]
	\centering
	\includegraphics[width=\columnwidth]{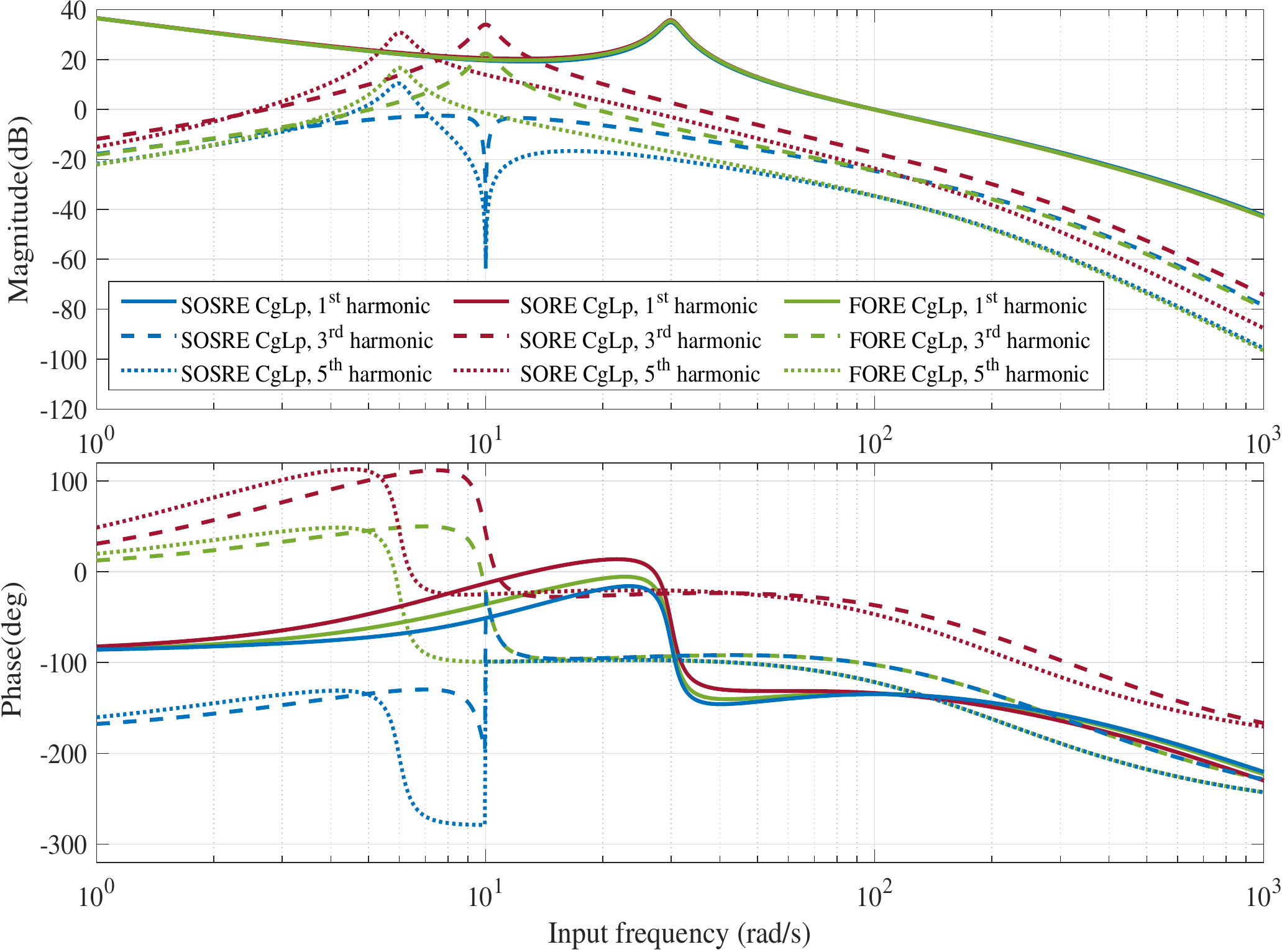}
	\caption{Comparison of the $1^{st}$, $3^{rd}$ and $5^{th}$ harmonic of open loop system using controllers designed based on CgLps introduced in~(\ref{eq:partial_SORE}),~(\ref{eq:conventional_SORE}) and~(\ref{eq:FORE}) in series with plant introduced in~(\ref{eq:plant}). All systems have matching $1^{st}$ order harmonic gain but significantly different $3^{rd}$ and $5^{th}$ harmonic gain.}
	\label{fig:openloop_hosidf}
\end{figure} 
For instance consider a mass-spring-damper system as:
\begin{equation}
\label{eq:plant}
P(s)=\frac{1}{11.11s^2+40s+10000}
\end{equation}
which has a resonance at $30~\text{rad}/{\text{s}}$ and is desired to be controlled with bandwidth of $100~\text{rad}/{\text{s}}$.
Following the instructions in~\cite{saikumar2019constant}, three controllers have been designed based on CgLps compared in Section~\ref{sec:comparison} in the framework of PID. The architecture of the designed controllers is depicted in Fig.~\ref{fig:openloop_blockdiagram}. The CgLps all have $\omega_{r\alpha }=10~\text{rad}/{\text{s}}$ which is one third of the plant's resonance, and all produce the same phase lead at the frequency of the bandwidth.\\
The overall quadratic stability of the closed loop reset system when the base linear system is stable can be examined by the following condition~\cite{guo2015analysis}.
\begin{thm}             
	There exists a constant $\beta \in \Re^{n_r\times 1}$ and positive definite matrix $P_\rho \in \Re^{n_r\times n_r}$, such that the restricted Lyapunov equation
	\begin{eqnarray}
	P > 0,\ &A_{cl}^TP + PA_{cl} &< 0\\
	&B_0^TP &= C_0
	\end{eqnarray}
	has a solution for $P$, where $C_0$ and $B_0$ are defined by
	\begin{align}
	C_0=\left[\begin{array}{ccc}
	\beta C_{p} & 0_{n_r \times n_{nr}} & P_\rho
	\end{array}\right] , & &  B_0=\left[\begin{array}{c}
	0_{n_{p} \times n_{r}}\\
	0_{n_{nr} \times n_{r}}\\
	I_{n_r}
	\end{array}\right].
	\end{align}
	And 
	\begin{equation}
		A_{\rho}^TP_\rho A_{\rho} - P_{\rho} \le 0
	\end{equation}
	$A_{cl}$ is the closed loop A-matrix.
	% 	\begin{equation}
	% 	A_{cl} =\left[\begin{array}{cc}
	% 	A_p & B_p C_r\\
	% 	-B_rC_p & A_r
	% 	\end{array}\right] 
	% 	\end{equation}
	% 	in which $(A_r,B_r,C_r,D_r)$ are the state space matrices of the controller defined by Eqn. \ref{eq:reset} with 
	$n_r$ is the number of states being reset and $n_{nr}$ being the number of non-resetting states and  ${n_{p}}$ is the number states for the plant.
	$A_p,B_p,C_p,D_p$ are the state space matrices of the plant.
\end{thm} 
This theorem requires the base linear system to stable. The weak tamed derivative which provides $5^\circ$ phase margin for the base linear system, exists to fulfil this requirement. Thus the overall controller phase margin for all CgLps is about $45^\circ$.\\ 
Figure~\ref{fig:openloop_hosidf} depicts the open loop HOSIDF. As expected, the third harmonic resonance happens at $10~\text{rad}/{\text{s}}$ and amplifies the third order harmonic for FORE CgLp and conventional SORE CgLp, while the notch-like behaviour of the SOSRE CgLp cancels the effect of the resonance peak. According to previous discussions, one can expect the SOSRE CgLp to have a better performance in terms of precision in the range of frequencies at which it has smaller third order harmonic. In particular, this will be the case at $10~\text{rad}/{\text{s}}$, where the other two CgLps have significantly larger third order harmonic.\\ 
Moreover, it can be seen in Fig.~\ref{fig:openloop_hosidf} that while the peaks of higher order harmonics are descending with respect to their order, the notch-like behaviour has also further decreased the peak of the $5^{th}$ harmonic for SOSRE. This also strengthens the expectation for better performance of SOSRE CgLp in terms of steady state tracking precision.\\
In order to validate the discussion, a simulation has been done using Simulink in Matlab. Its results are presented in the following section.\\ 
\begin{figure}[t!]
	\centering
	\scalebox{0.9}[0.9]{
		\includegraphics[width=\columnwidth]{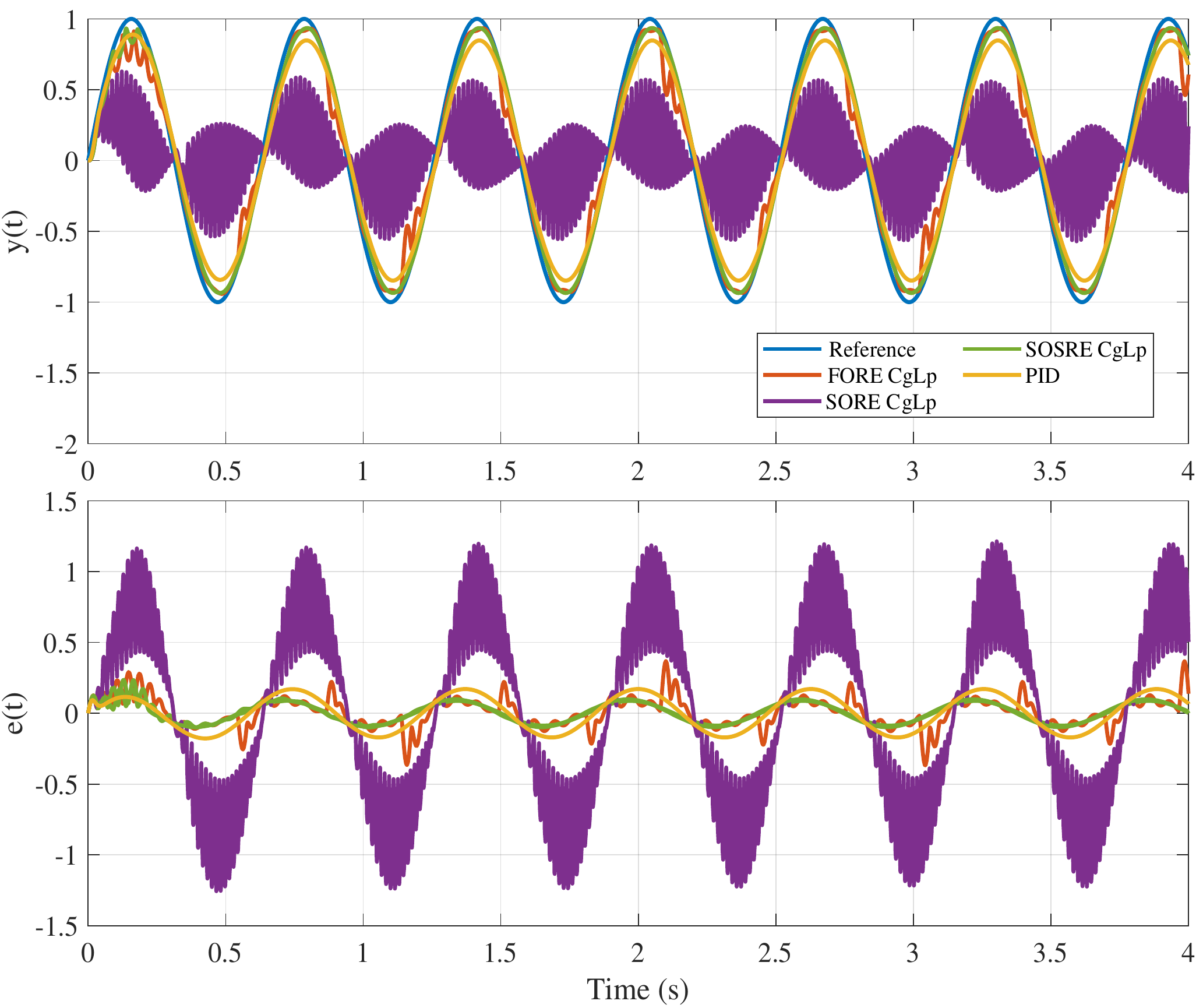}
	}
	\caption{Simulation results of output, $y(t)$, and error, $e(t)$, based on architecture presented on Fig.~\ref{fig:openloop_blockdiagram}. Reference is $r(t)=\sin(10t)$.}
	\label{fig:sim_results_error}
\end{figure}
%\begin{remark}
%	It is very important to notice that HOSIDF analysis is based on a sinusoidal input to the system; therefore, the notch behaviour will only exist if the steady state input to the reset element is periodic. For the case of a closed loop system, this requires the base linear system to be stable.
%\end{remark}	
%Since a tamed derivate is a linear filter, its output is periodic, provided that its input (i.e., $e(t)$) also is. Assuming $r(t)=\sin(\omega t)$, in order to make sure that the steady state part of the $e(t)$ is periodic, it is trivial that the base linear system should be stable.
\section{Simulation results}
\begin{figure}[t!]
	\centering
	\includegraphics[width=\columnwidth]{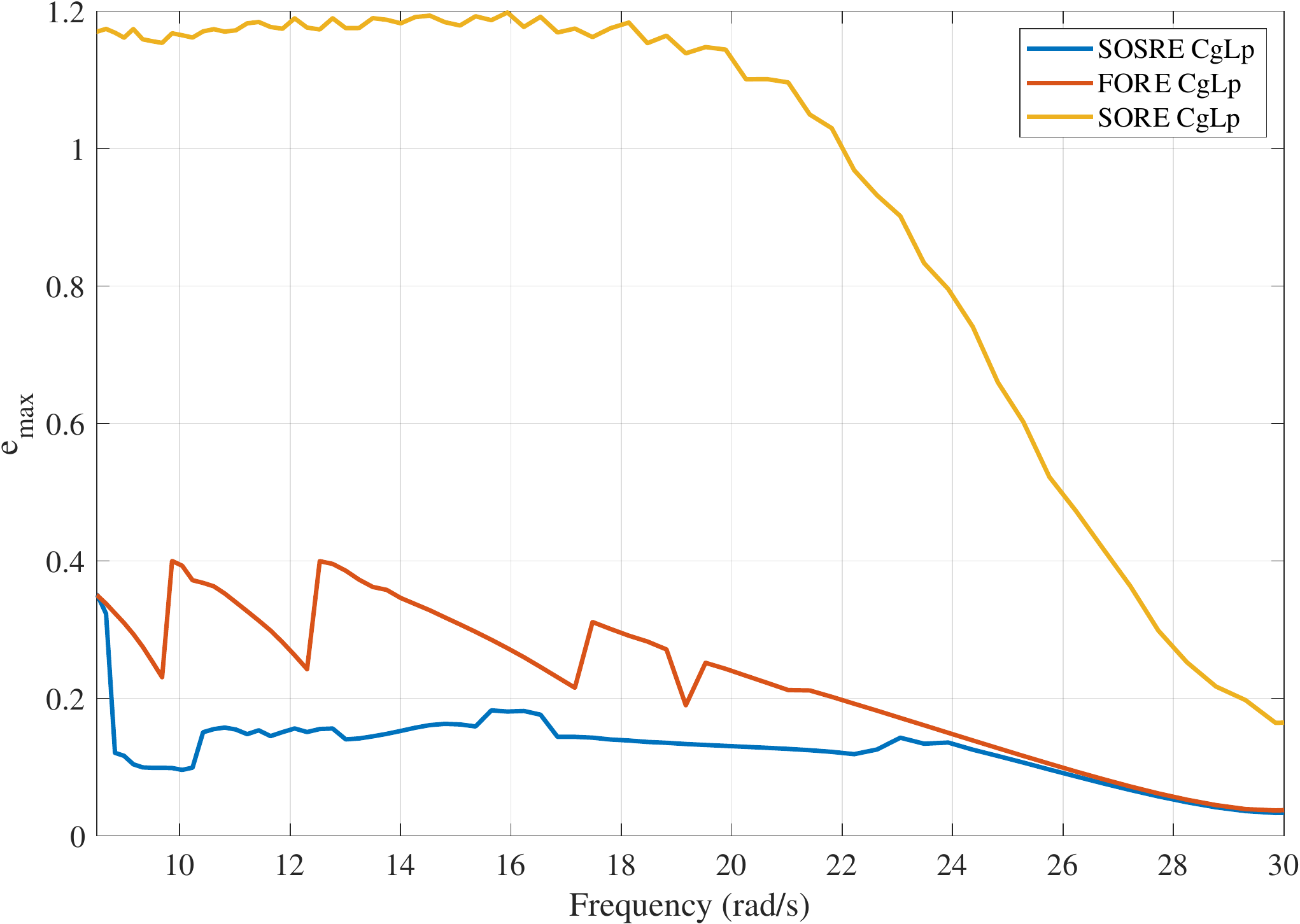}
	\caption{The steady state maximum error of three CgLp elements for sinusoidal input.}
	\label{fig:compare_range_freq}
\end{figure}

To validate the hypothesis in the time domain, and the improvements observed in the frequency domain, and also in order to be able to compare the controllers in terms of precision, a simulation has been done for a sinusoidal input with frequency of $10~\text{rad}/{\text{s}}$. Furthermore, for the sake of completeness, results are also obtained and compared with a linear PID, in which $\omega_i=10~\text{rad}/{\text{s}}$, $\omega_d=26.3~\text{rad}/{\text{s}}$ and $\omega_t=380~\text{rad}/{\text{s}}$ and there is a second order low pass filter with same characteristics as there is in CgLps. It should be noted that PID is tuned in such a manner that it provides the same phase margin as other controllers.  \\
\begin{table}[t!]
	\centering
	\caption{Comparison of $L_2$ and $L_{\infty}$ of the steady state error of each controller.}
	\label{tab:errors}
	\begin{tabular}{@{}lll@{}}
		\toprule
		Controller & $L_2$ & $L_\infty$ \\ \midrule
		SOSRE CgLp      & 0.099 & 0.069      \\
		Conventional PID        & 0.171 & 0.123      \\
		FORE CgLp     & 0.368 & 0.105      \\
		SORE CgLp     & 1.214 & 0.672      \\ \bottomrule
	\end{tabular}
\end{table}
\begin{figure}[t!]
	\scalebox{0.9}[0.8]{
		\centering
		\includegraphics[width=\columnwidth]{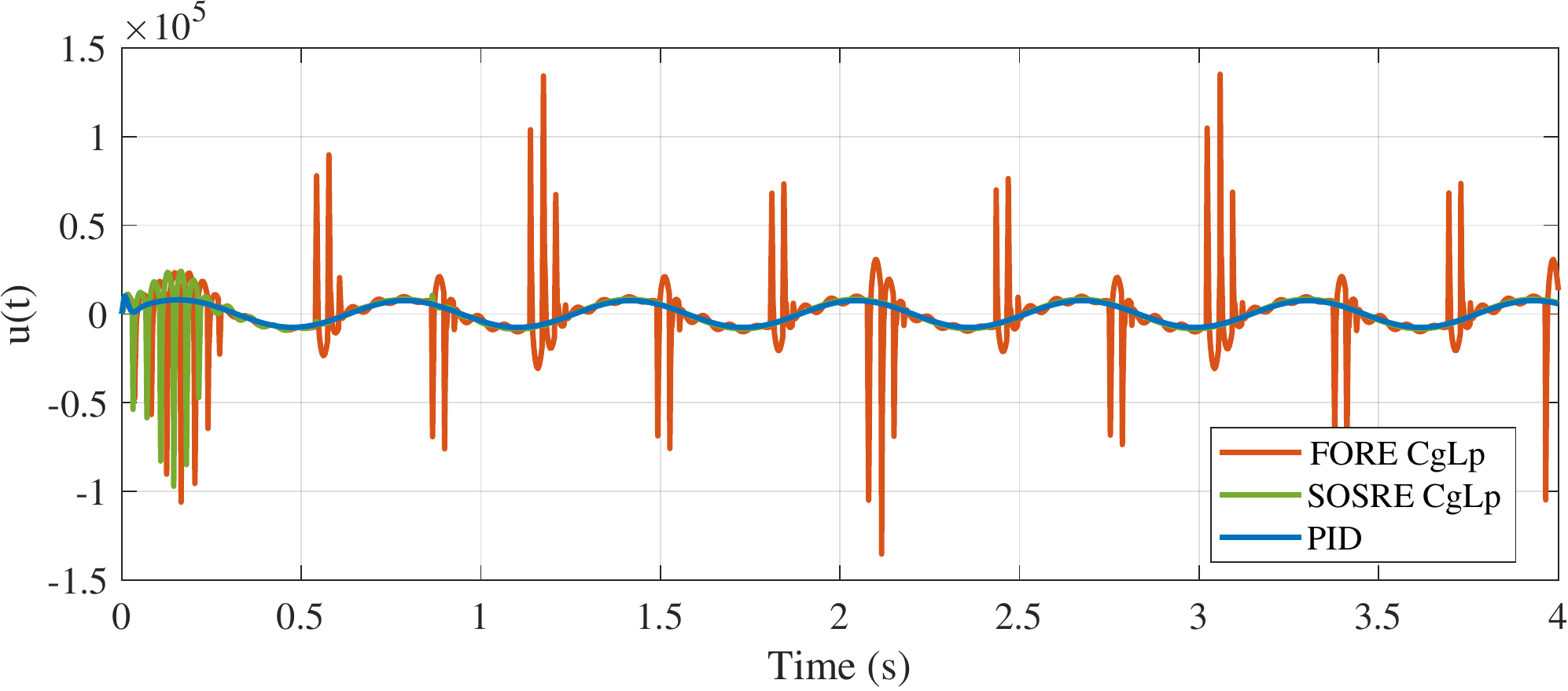}
	}
	\caption{Control input comparison of FORE CgLp, SOSRE CgLp and PID.}
	\label{fig:sim_res_cont_out}
\end{figure}
\begin{figure}[t!]
	\centering
	\scalebox{0.95}[0.9]{
		\includegraphics[width=\columnwidth]{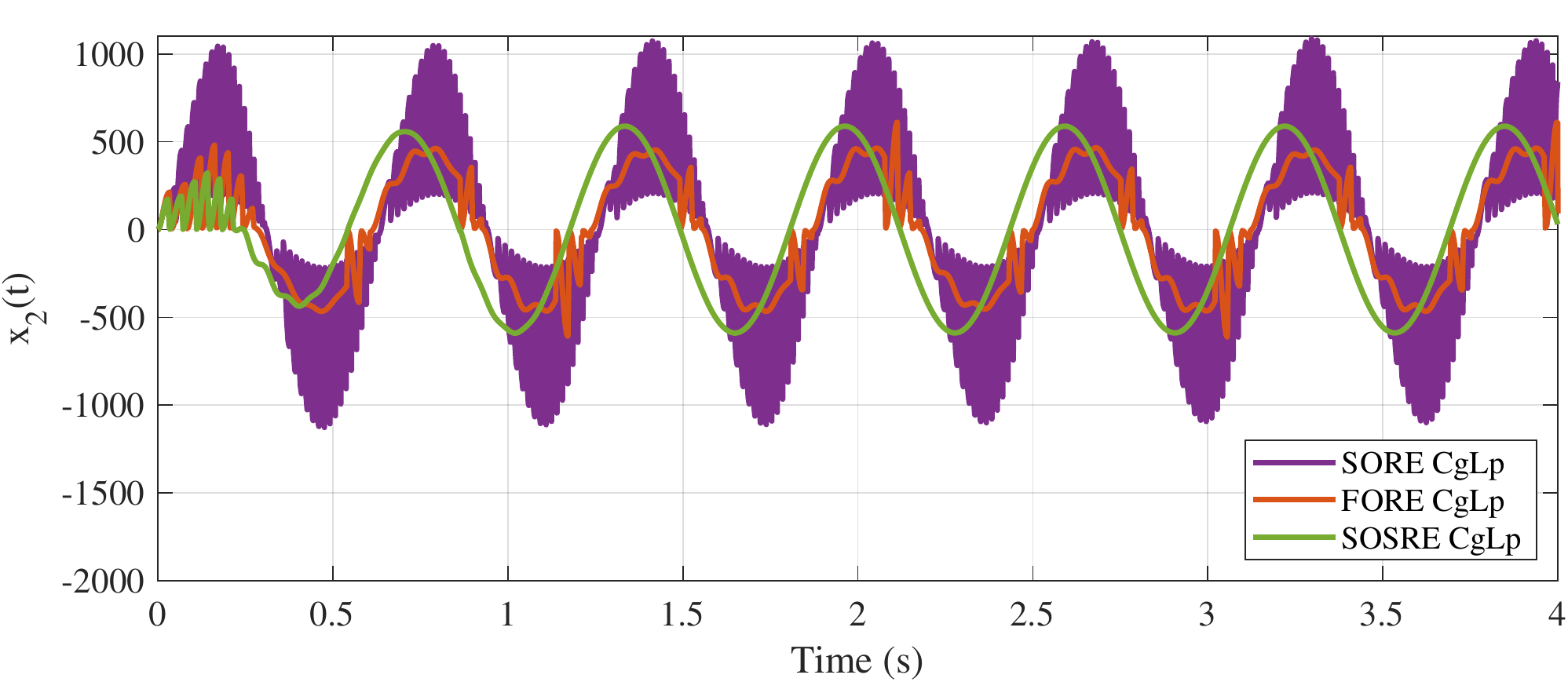}
	}
	\caption{Simulation result of $x_2(t)$ for all three CgLps.}
	\label{fig:sim_res_x1}
\end{figure}
The output and error for each controller is presented in Fig.~\ref{fig:sim_results_error}. Table~\ref{tab:errors} represents the RMS ($L_2$) and maximum ($L_{\infty}$) of the steady state error of each controller. One can readily observe that conventional SORE CgLp has the poorest performance and SOSRE CgLp outperforms the other three controllers by nearly an order of magnitude improvement in precision. Simulation results clearly validates the better performance of proposed SOSRE CgLp in terms of steady state tracking precision for frequency of the notch-like behaviour. It worth mentioning that the estimation of $L_2$ and $L_{\infty}$ of closed loop steady state error based on DF of the three CgLps are 0.099 and 0.069, which is the same as the SOSRE CgLp. Hence showing that at this frequency the minimization of harmonics makes the DF completely reliable.\\% Since the magnitude of higher order harmonics for SOSRE CgLp is not higher than the other two at any frequency, its performance is expected to be slightly better in all frequencies; however it should be subject to further investigations.  \\
Reset-based controllers usually have large peaks in their output and thus are not very control effort efficient. Another characteristic of SOSRE CgLp is a relatively small control input at the frequency of notch-like behaviour which is almost comparable with PID. Since conventional SORE CgLp has too poor a performance in terms of accuracy and has 2 orders of magnitude larger $u(t)$, Fig.~\ref{fig:sim_res_cont_out} only depicts the comparison between FORE CgLp, SOSRE CgLp and PID which validates the claim.\\
The simulations results also validate the claim of Remark~\ref{remark}. Figure~\ref{fig:sim_res_x1} depicts the value of $x_2(t)$, introduced in Fig.~\ref{fig:GSORE_blockdiag} for all three CgLps. One can observe that after transient response, no reset is seen for the $x_2$ state of the SOSRE CgLp.\\
In order to have a clearer view of the higher order harmonic effect on the steady state tracking error, a further investigation has been carried out on other frequencies around the frequency of SOSRE higher order harmonic notch. As illustrated in Fig.~\ref{fig:compare_range_freq}, the $L_{\infty}$ of the steady state error of the SOSRE CgLp deviates from that of the FORE CgLp from $8~\text{rad}/{\text{s}}$ till $30~\text{rad}/{\text{s}}$. It shows that higher order harmonic notch-like behaviour of the SOSRE also improves the performance not only at the frequency of the notch itself, but also at frequencies around. However, a complete closed-loop performance analysis of these elements is subject to further investigation.\\
For this example, one may suggest using a notch filter to cancel out the resonance of the plant for cancelling the corresponding peaks in higher order harmonics. Such a filter will remove the free gain available from the resonance in first order harmonic. However using SOSRE CgLp one can reduce higher order harmonics without changing the first order one. Furthermore, higher order harmonics have their adverse effect in frequencies other than their peaks and wherever the higher order harmonic notch of the SOSRE is tuned to be, e.g., a critical working frequency of the system, the performance is guaranteed to be the same as DF.  
\section{Conclusion}

This paper studied a special case of a SORE CgLp, in which only one state is being reset. It was shown that when input and output of a reset element's base linear system have the same phase at certain frequencies, the reset action will be of no effect and the element will behave like a linear one at the same frequencies. In the special architecture of SOSRE CgLp presented in this paper, based on the aforementioned fact, a notch-like behaviour in higher order harmonics gain is found and at the same time the first order harmonic gain behaviour is conserved.\\
 In this paper, the notch-like behaviour was used to cancel out the resonance peak of the third harmonic of the system. However, the application is not restricted to this, and wherever the higher order notch is placed, the performance is guaranteed to be the same as DF. The simulation results validated the claim that the SOSRE CgLp is behaving linear in terms of steady state output at the frequency of higher order notch and also has better performance in terms of precision at a range of frequency around it.\\
As ongoing works, a complete closed loop analysis will be carried out on this element; moreover, its behaviour will be investigated in presence of noise and disturbance. Furthermore, the simulation results will be validated in practice. 

%%%%%%%%%%%%%%%%%%%%%%%%%%%%%%%%%%%%%%%%%%%%%%%%%%%%%%%%%%%%%%%%%%%%%%%%%%%%%%%%

%%%%%%%%%%%%%%%%%%%%%%%%%%%%%%%%%%%%%%%%%%%%%%%%%%%%%%%%%%%%%%%%%%%%%%%%%%%%%%%%

%%%%%%%%%%%%%%%%%%%%%%%%%%%%%%%%%%%%%%%%%%%%%%%%%%%%%%%%%%%%%%%%%%%%%%%%%%%%%%%%
%\section*{APPENDIX}
%\section*{ACKNOWLEDGMENT}
%
%The preferred spelling of the word ÒacknowledgmentÓ in America is without an ÒeÓ after the ÒgÓ. Avoid the stilted expression, ÒOne of us (R. B. G.) thanks . . .Ó  Instead, try ÒR. B. G. thanksÓ. Put sponsor acknowledgments in the unnumbered footnote on the first page.

%%%%%%%%%%%%%%%%%%%%%%%%%%%%%%%%%%%%%%%%%%%%%%%%%%%%%%%%%%%%%%%%%%%%%%%%%%%%%%%%

\bibliography{ifacconf}

\end{document}